# A bibliometric approach to Systematic Mapping Studies: The case of the evolution and perspectives of community detection in complex networks

**Camelia Muñoz-Caro***. SciCom Research Group. Escuela Superior de Informática. Universidad de Castilla-La Mancha. Paseo de la Universidad 4, 13004 Ciudad Real, Spain. Email: camelia.munoz@uclm.es

**Alfonso Niño**. SciCom Research Group. Escuela Superior de Informática. Universidad de Castilla-La Mancha. Paseo de la Universidad 4, 13004 Ciudad Real, Spain. Email: alfonso.nino@uclm.es

**Sebastián Reyes**. SciCom Research Group. Escuela Superior de Informática. Universidad de Castilla-La Mancha. Paseo de la Universidad 4, 13004 Ciudad Real, Spain. Email: sebastian.reyes@uclm.es

*Corresponding author.

## Acknowledgements

The authors wish to thank the Consejería de Educación y Ciencia de la Junta de Comunidades de Castilla-La Mancha (grant # PEII-2014-020-A). The economic support of the Universidad de Castilla-La Mancha is also acknowledged.

# Abstract

Critical analysis of the state of the art is a necessary task when identifying new research lines worthwhile to pursue. To such an end, all the available work related to the field of interest must be taken into account. The key point is how to organize, analyze, and make sense of the huge amount of scientific literature available today on any topic. To tackle this problem, we present here a bibliometric approach to Systematic Mapping Studies (SMS). Thus, a modify SMS protocol is used relying on the scientific references metadata to extract, process and interpret the wealth of information contained in nowadays research literature. As a test case, the procedure is applied to determine the current state and perspectives of community detection in complex networks. Our results show that community detection is a still active, far from exhausted, in development, field. In addition, we find that, by far, the most exploited methods are those related to determining hierarchical community structures. On the other hand, the results show that fuzzy clustering techniques, despite their interest, are underdeveloped as well as the adaptation of existing algorithms to parallel or, more specifically, distributed, computational systems.

# Introduction

New research lines worthwhile to pursue by the scientific and technological community must be properly oriented. To such an end, all the available work related to the field of interest must be taken into account. The main problem is how to organize, analyze, and make sense of the huge amount of scientific literature existing today on any topic. As an example of the difficulties faced when identifying a new research topic, we have the case of community detection in complex networks.

The complex networks paradigm represents a powerful way to describe and handle complex systems. Under this paradigm, a system is described as a set of nodes, the entities of the system, interrelated by a series of edges representing the interactions among the entities. The interest in this approach was triggered by the seminal works of Watts and Strogatz on small-world networks (Watts & Strogatz, 1998), and Barabási and Albert on the generation of scale-free networks by preferential attachment (Barabási & Albert, 1999). Since then, the field has grown rapidly, being applied to the interpretation and prediction of the behavior of communication, social, biological, epidemiological, chemical or transportation networks (Costa et al., 2011).

Many different studies can be carried out on networks (Newman, 2010; Barabási, 2016), but one of the most interesting is community detection. Community structure, or clustering, is a consequence of inhomogeneities in the network. Thus, some nodes are more "strongly" connected among them than with others. This fact defines communities of nodes (Fortunato, 2010) with probably similar state or behavior. The problem is how to characterize this "strength" of the connection. No single approach exists; it usually depends on the problem considered, although it is frequently related to edge density (Fortunato, 2010). Therefore, many different types of methods exist such as hierarchical, partitional or spectral algorithms, among others (Schaeffer, 2007; Fortunato, 2010; Coscia et al., 2012; Xie et al., 2013; Barabási, 2016).

Community detection in complex networks presents a great interest due to the huge number of practical applications. The key point is that nodes in the same community represent entities with similar characteristics. So, in a social network, communities represent interrelated individuals such as families, friend circles, or co-workers. In communications networks, they correspond to systems strongly interrelated where ease of communication would represent an increase of the total efficiency of the system. In a sales system, communities can represent sets of customers with similar preferences.

Being such an interesting field, with so many potential applications, it has attracted a lot of attention. However, to develop new contributions useful to the scientific and technological community new studies must be properly oriented. So, what is the real scope of the current studies in the field? What aspects are over or understudied? What are the most, or least, used community detection methods? These questions are important to identify new research lines and programs. The problem is how to answer systematically these questions using the wealth of information in the available research literature on the topic.

As a research area evolves, the number of available reports and results providing new knowledge, primary studies, increases. Therefore, it becomes important to summarize and overview all or part of the area, secondary studies. The traditional approach to secondary studies is the Literature Review, LR. In an LR, the goal is to summarize, interpret, evaluate, and organize the information contained in the primary studies on a given topic. LRs are well-known in any scientific field, but no specific formalization of the review process does exist. With the aim to reduce bias and provide a clear understanding on a given subject, a different approach was introduced in evidence-based medicine (Cook et al., 1992): The Systematic (Literature) Review, SLR. An SLR follows a standardized protocol to reduce bias, identify relevant literature, assess its significance, and summarize it. In this form, we can answer specific questions on a given research area (Khan et al., 2003). SLRs synthesize the results of multiple independent studies in a single study. Thus, in current health care, SLRs are considered one of the highest levels of evidence available for assessing the validity and effectiveness of treatments, tests, and biochemical associations (Belter, 2015; DiCenso, Bayley & Haynes, 2009). The importance given to SLRs have led to recommended standards for systematic reviews such as those proposed in (National Research Council, 2011; Shamseer et al., 2015). In addition, collaborative groups have been created to develop freely available SLRs on different topics such as the Cochrane (Cochrane, 2016) and Campbell (Campbell Collaboration, 2016) collaboration groups. The SLR approach has been adapted to several fields such as Sociology (Petticrew & Roberts, 2005) and, lately, Software Engineering (Kitchenham et al., 2004; Dybå et al., 2005; Jørgensen et al., 2005; Kitchenham & Charters, 2007; Kitchenham et al., 2010; Wohlin et al., 2013). In this last technological field, the SLR approach has proven to be very successful. However, SLRs are a hard and time-consuming activity since a proper consideration of all the evidence on a given topic needs literature searches close to 100% and an analysis of the references to determine their relevance for the study (Belter, 2015).

An SLR provides a concrete, detailed, deep, view of the research done in an area. In fact, an SLR aims to answer a specific research question such as, is method A more time efficient than method B? However, for studies trying to organize and classify primary studies a more specific form of SLR has arisen, the Systematic Mapping Study, SMS (Petticrew & Roberts, 2005; Kitchenham & Charters, 2007; Petersen et al., 2008). An SMS follows a protocol similar to an SLR, but the goal is to get a general picture of the topic of interest. Thus, an SMS maps out a research area using different classifications of primary studies. The SMS is especially well suited to identify trends by analyzing the frequency distribution of publications and the time evolution of specific aspects of the topic considered. So, SMSs are very useful to identify holes in current studies and to find topics that represent opportunities for new research (Kitchenham et al., 2011). Specific guidelines for performing SMSs have been proposed (Petersen et al., 2008; Petersen et al., 2015). In the studies so far, SMSs are usually focused in very specific topics and, therefore, the relatively small number of references to handle allows for a "by hand" selection and labeling of reports and papers. However, if we want to take advantage of the vast amount of information contained nowadays in the scientific literature, we need to adopt a different approach.

When the volume of primary studies to process is very large, we can apply a bibliometric approximation. First introduced by Pritchard (1969), bibliometrics is defined as the statistical analysis of bibliographical information. In short, it is a way to analyze scientific publications quantitatively (De Bellis, 2009). Therefore, statistics can be used to characterize the large sets of primary studies involved in SMSs studies rather than an individual analysis of every single reference.

In this work, we present a bibliometric approach to SMSs to determine the current state of a given research field. To such an end, we define an adapted SMS protocol, relying on the metadata associated to primary studies, to analyze and interpret the information contained in the large amount of references available today in scientific literature. This approach is applied to determine the current state and research opportunities of community detection in complex networks. The study focuses in quantifying the evolution of the field as well as in determining the relative weight of the different community detection methods and computational models used.

# Methodology

Systematic studies are applied through a previously defined protocol. To define a protocol, the starting point is to consider that any systematic study is developed in five standard steps (Khan et al., 2003; Kitchenham et al., 2004; Kitchenham & Charters, 2007): identify the questions we want to answer; locate the associated primary studies; ensure the quality of the studies; extract the evidences contained in the studies; organize and interpret the results. It is important to note that the whole process is oriented to answer specific research questions. So, they must be carefully selected.

The protocol presented in this work takes into account the guidelines defined in (Petersen at al., 2008), adapting the general SMS scheme in the way shown in Figure 1. Here, we organize the five basic steps in two levels. The first one, outer part of the diagram, represents the action implementing the step. In the inner part of the diagram, we have the outcome obtained after applying the previous action. The different steps are defined as follows.

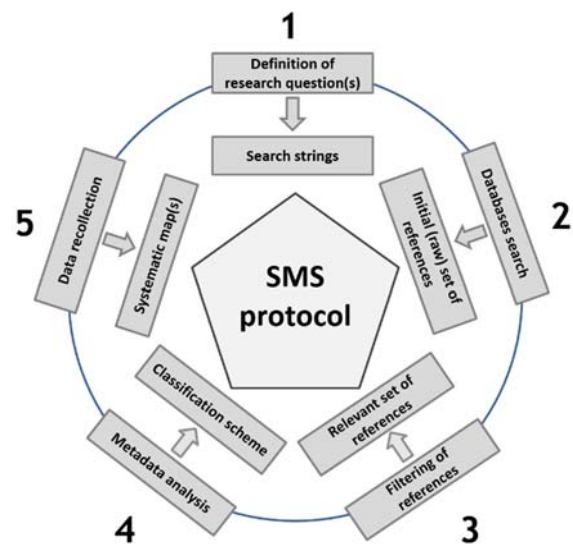

**FIG. 1** Systematic Mapping Study (SMS) protocol proposed and used in this work.

*First step*

In the first level of the first step the research questions we want to answer are defined. These are specific statements addressing unambiguously the aspects of the topic we are interested in (Khan et al., 2003; Kitchenham et al., 2004; Kitchenham & Charters, 2007).

Research questions can be built focusing in one or more of the concepts of population, intervention, comparison, outcome, context and experimental design associated to the topic researched, as proposed in (Kitchenham & Charters, 2007). To make them operative, each research question is associated with an appropriate search string that can be used in scientific literature databases. This defines the second level of the first step. Thus, we select keywords representing, as broadly as possible, the core structure of the topic under research. Logical operators are used to concatenate the keywords in such a way that they correspond to the associated research question.

*Second step*

In the second step, in its first level, we select and search the scientific databases covering all the desired aspects of the research questions, for instance, databases including or not technological or humanities content. Furthermore, the time interval and metadata fields where the search is going to be carried out must be defined. Finally, the sources of research (jobs, papers, collective books, conferences) are also determined. The databases are searched using the search strings previously defined in the first step. This results in an initial, or raw, set of references, as shown in the second level of the current step. Actually, we obtain a set of references for each of the used databases. Except if the scope of the study is very narrow, we obtain several thousand references. On one hand, this represents a useful source of information. On the other, however, it is clear the need to automate the processing of such large a dataset.

*Third step*

The raw set of references resulting from the search on the different databases usually presents inconsistencies. These are due to the presence of errors in the databases referencing process, to the use of the same search string on different databases, and to the fact that no search string is totally free from ambiguity. Thus, it is necessary a work of integration, filtering and cleaning of the initial references set, first level of this step, before drawing any conclusion on the topic of interest. Actually, this is nothing but a quality assurance process.

In a usual SMS, the number of references is typically small. For instance, in a tertiary study about systematic studies in the software engineering field (Kitchenham et al., 2010), only three of the 33 studies considered deal with more than 155 references. With small sized reference sets, any kind of revision process can be done by direct inspection of all the individual papers in the set. In our case, with typically thousands of references in the raw set, this is an unfeasible approach. Thus, we resort to the use of appropriate bibliographic management tools such as Mendeley (Mendeley, 2016), Zotero (Zotero, 2016) or Thomson Reuters' EndNote (EndNote, 2016), and to the definition of a case adapted protocol based on the metadata available from the citation databases searches. In this form, we identify the possible presence of three typical problems,

- *Duplicate references:* Duplicate publications can affect the reliability of meta analyses, as the same dataset might be counted more than once (Gurevitch & Hedges, 1999; Tramer et al., 1997; Wood, 2008). Due to the large number of references in these studies, the problem can be addressed merging the sets of references through an appropriate configuration of the eliminate duplicates capabilities of reference management systems, see (Mendeley, 2016; Zotero, 2016; EndNote, 2016).

- *References without authors:* This problem is simple to correct just by sorting the references by author. Authorless references appear as a single set, which allows their direct removal.

- *References addressing topics unrelated to the topic of interest:* Unrelated references or false positives are a consequence of ambiguity in the search keywords, and they need to be removed. Since individual analysis of each reference is unfeasible, we propose to sample the dataset using a random selection procedure. Thus, we select randomly a first reference, analyzing the title, abstract, and keywords to determine the nature of the work. If the work is unrelated to the topic of interest, we characterize it using the "keywording" procedure described in (Petersen et al., 2008). Keywording is a systematic way to characterize a set of existing papers. In particular, the topic, keywords, concepts and context associated to the current paper are registered. Then, we continue randomly selecting papers, repeating the procedure until no additional false positives are found after ten new randomly selected references. After that, similar false positives are considered together, forming a set. Thus, using the information registered for each reference, a group of keywords is generated describing the nature and contributions of the jobs in the set. We cannot assume that all papers belonging to the different sets can be identified in this form. However, this is a good way to identify, through the sets, the types of unrelated references. Then, appropriate search strings are built representing each set of unrelated references. With this information, we search the whole dataset to locate and remove all the references unrelated to the topic.

The previous activities do not ensure, though, that the resulting references dataset is free from false positives. Since the size of the dataset is large (typically several thousand references) direct analysis of each one is not a feasible solution. Thus, to have an objective indication of the final dataset quality, we resort to standard statistical techniques. We propose to select a representative random sample from the relevant references dataset, individually analyzing the associated references. Using typical statistical values (Moore et al., 2014), we determine the sample size providing a confidence level of 95% with a confidence interval of 10% from the whole references population. These references are randomly selected and considered to determine the quality of the reference filtering process.

As shown in Figure 1, applying the previous considerations to the raw set of references, we identify, in the second level of this step, the meaningful set of studies. In this form, we obtain the processed, relevant, set of references.

*Fourth step*

In the fourth step, see Figure 1, we determine how to obtain the data needed to answer the research questions. For that, a two steps process is proposed:

- First, we determine what information is needed to answer the research questions. Typically, the information corresponds to frequencies of publications or to the number of publications dealing with specific aspects of the topic considered. In addition, the metadata fields where the information is to be located (Title, Abstract, Keywords…) are also selected.
- Second, we choose the keywords, and logical relationships between them, which would lead to the desired information.

The result of this activity, second level of this step, is a set of qualified search strings defining specific searches on the relevant references set.

*Fifth step*

In the last step, we apply the search strings corresponding to the qualified metadata terms, obtained in the previous step, collecting the data associated. The results are graphically represented, analyzed and interpreted, second level of this step, providing the information we need to answer the research questions.

## Application

To test the defined protocol, see Figure 1, we consider as case study the current state and perspectives of community detection in complex networks.

*Step 1. Definition of research questions*

The starting point of the SMS is a clear definition of the study scope. To such an end, we need to state one or several research questions. In the present case, our focus is two- (three) fold: the evolution and current state of community detection in complex network and the relative weight of the different methods and computational models available today. Thus, we give concrete form to this broad question as three specific research questions, RQs. The RQs defined here are,

**Research question 1 (RQ1).** *What is the evolution of community detection studies in complex networks?*

**Research question 2 (RQ2).** *What is the relative weight of the different community detection methods?*

**Research question 3 (RQ3).** *What is the relative weight of the different computational models used in community detection methods/studies?*

To answer these questions, we need to define an appropriate search string incorporating all the aspects present in them. To define this string as broadly as possible, we consider the different definitions of our problem. In this case, we select community detection (or clustering) methods (or algorithms) in complex networks (or graphs). The search string, making use of wild characters, quotation marks for literal expressions, logical operators, and parenthesis to express operator precedence is:

*("communit\* detection" OR "graph clustering" OR "communit\* structure") AND (algorithm\* OR method\*) AND (network\* OR graph\*)*

In the next step, we search the scientific literature for references corresponding to the selected search string.

*Step 2. Databases search*

To cover the maximum possible spectrum of primary sources in the arts, humanities, science and engineering literature, as close to 100% as possible, we use several scientific databases, in particular:

- Elsevier's Scopus

- Thomson Reuters' Web of Science

- IEEEXplore

- Elsevier's ScienceDirect

We include Scopus and Web of Science to take advantage of their strong, but not identical, coverage of different fields. In particular, Scopus provides a larger coverage of publications, whereas Web of Science allows to include old literature on the topic because it indexes works since 1900. On the other hand, IEEEXplore is included to account for engineering and computer science literature not included in the previous databases. Finally, ScienceDirect is selected because it offers specific coverage of relevant scientific literature, being a complement of its sister database Scopus.

For each database, we select the full range of available years. The search was carried out on April 26, 2016. We apply the search string on the title, abstract and keywords of the sources. Finally, we select all possible sources (journals, conferences, collective books…) for the queries.

The number of sources retrieved after searching each database is collected in Table 1. We can see that 9160 references are recovered from the different databases in a period ranging from 1985 to 2016. Since 2016 is not yet over at the time of writing this paper, data for this year are incomplete. The next step is to integrate and filter the whole set of references.

**TABLE 1**. Result of the searches in the different citation databases considered in this work.

| Database | References |
|---|---|
| Scopus | 4297 |
| Web of Science | 3030 |
| IEEE Xplore | 1295 |
| Science Direct | 537 |
| *Total references* | 9160 |
| *Timespan* | 1985-2016 (partial) |

*Step 3. Filtering of references*

In this step we select the relevant references from the results of the searches in the different databases. Essentially we have here a process of integration and filtering.

By performing an initial, random, inspection of the raw references in the raw set, we identify the existence of the three problems described in the Methodology section,

• Duplicated references

• References without authors

• References addressing topics unrelated to the topic of interest

To solve the first problem, we use Thompson Reuters' EndNote reference manager (EndNote, 2016) to eliminate duplicates. Then, we first compare references based on title, year, volume, and journal. This very specific search serves to clean the bulk of duplicated references. Additional inspection of the dataset reveals still some duplications arising from discrepancies in journal or conference name, volume, pages or years in the different databases used. Therefore, after having applied the first filter, we search again for duplicated references based only in title. This strategy seems to remove most, if not all, the remaining duplications. In this form, the number of references is reduced to 5243.

The second problem is corrected sorting the references dataset by author. Thus, the authorless references appear as an independent set. We check each of these references individually founding that they correspond to introductions or presentations of collective and proceedings books. After removal, the final dataset contains 5223 different elements.

Finally, the third problem is addressed using the "keywording" technique described previously. In this form, we identify three types of references wrongly included in the dataset (false positives):

- References dealing with community structure in ecosystems (mainly marine), without relation to community detection methods or algorithms.
- Unrelated references dealing with neural networks.
- References on graphics and climate without any relationship to community detection.

The "keywording" procedure provides a set of keywords associated to the false positives. Therefore, we extract and inspect all the references in the dataset containing, independently, the following terms: *microbial\*, ecosystem\*, ecologic\*, bacterial, marine, fish\*, sponge\*, plant\*, parasit\*, neural network\*, climate* and *graphical\**. The resulting references are individually analyzed. All references unrelated to the topic of this study were removed. The final set results in 4846 references.

To have an objective index of the dataset quality, we select a representative sample from our 4846 references dataset. As previously indicated, we determine the sample size needed to attain a confidence level of 95% with a confidence interval of 10%. The corresponding sample size is 94 references (Moore et al., 2014). Hence, we randomly choose 94 references from the dataset, individually checking their metadata. All they actually deal with appropriate aspects of community detection. Therefore, on statistical grounds, we can be 95% certain that a minimum of 90% of our dataset can be considered free from false positives.

*Steps 4 and 5. Metadata analysis, data recollection and mapping*

For the sake of clarity these two last steps of the protocol, see Figure 1, are organized by research question.

**RQ1.** *What is the evolution of community detection studies in complex networks?*

To answer this question, we use the references dataset to determine the number of papers published per year in the available range. To such an end, we wrote a program to read the complete set of references, exported in RIS format, and determine the number of references per year. The results are collected in Figure 2, where we have excluded year 2016 since only partial data are available. We can observe that the references span a range of 30 years, from 1985 to 2015. As for the time evolution, Figure 2 shows a relative flat zone between 1985 and 2001 (no more than five references per year) followed by a steady and fast increase since 2002.

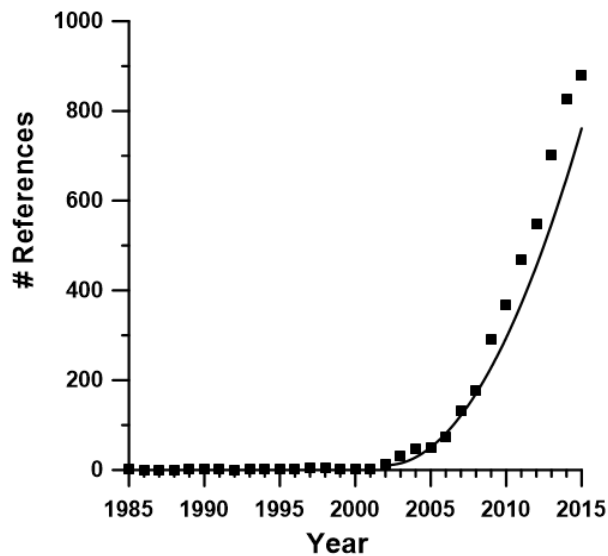

**FIG. 2** Number of references on community detection in complex networks as a function of time. The squares indicate experimental data. The continuous line is the regression function detailed in the text.

These findings can be interpreted by resorting to Rogers' diffusion of innovations theory originally presented in 1962. This theory describes how new ideas, practices or technologies spread into a social system (Rogers, 2003). In this context, innovation diffusion is a general process, not bound by the type of innovation studied, such that the process through which an innovation becomes diffused has universal applications to all fields developing innovations (Rogers, 2004). Diffusion of innovations theory has spread to many different fields, and thousands of studies support it (Rogers, 2004; Rogers, 2003). According to Rogers (2003), the variation of the number of adopters or users of a new idea over time follows a logistic (sigmoidal) function. This function exhibits a relatively flat first section followed by a fast increase leading to an inflection point. After that, the growing rate with time decreases until the curve reaches an upper relatively flat zone. This marks the maturity and maximum adoption of the idea. In the present study, community detection is the considered idea, and its adoption is measured by the variation with time of the number of references on the topic.

To analyze the variation of the number of references with time, we use the following approach. We assume that the number of references, $n_r$, depends on time, t, and we expand $n_r$ in a Taylor series around a given point, $t_0$, see equation (1).

$$n_r = c_0 + \sum_{l=1}^{\infty} \frac{1}{l!} \left(\frac{d^l n_r}{dt^l}\right)_{t_0} (t - t_0)^l = c_0 + c_1 (t - t_o) + c_2 (t - t_o)^2 + \cdots \quad (1)$$

This approach has been successfully applied to different domains such as the modelling of kinetic and potential functions in molecular vibrational hamiltonians (Liu et al., 1996), the development of job scheduling algorithms in distributed systems (Díaz et al., 2009), and the analysis of degree correlation variations in complex networks (Niño and Muñoz-Caro, 2013). Now, a multilinear regression can provide the value of the $c_i$ coefficients in equation (1) as well as additional useful statistics.

For applying equation (1), we use the data between years 2002 and 2015, expanding the series around year 2002. For the regression, we limit the polynomial to fourth degree and apply the stepwise selection procedure to include only the most statistically significant terms (Pekoz, 2015). The resulting polynomial is shown in equation (2),

$$n_r = 10.51 + 5.44 (t - 2002)^2 \quad (2)$$

with a coefficient of determination $r^2 = 0.994$ and standard deviation $\sigma = 24.74$ references.

On one hand, the excellent fit, with only one quadratic term, is an indication of the homogeneity of the dataset. On the other, the observed positive and growing slope of the number of references with time, see equation (2), indicates that community detection studies have not yet reached their inflection point. Thus, on the grounds of Rogers's theory (Rogers, 2003), there is still a growing interest in the field, which, therefore, seems to be far from exhausted.

The cause of the trend change observed Figure 2 can be determined by analyzing the years previous to 2003. As a global index for taking into account the time evolution of the field, we consider the cumulative number of citations per year. Thus, for each year, we use the aggregated value obtained by adding all the citations available for the references published that year. The results are collected in Figure 3. We can observe that until 2001 the cumulative number of citations is small, with a maximum of 74 for 1998. However, 2002 increases dramatically the count up to 5110 for 12 publications. Clearly, some 2002 publications triggered a huge interest in the field. By extracting the number of citations of each reference, the metadata allows identifying the cause of this fact. The reason is the publication of the papers by Girvan and Newman (2002), 3203 citations, and Ravasz et al. (2002), 1734 citations, which can be considered the starting point of the modern period of community detection studies (Barabási, 2016). Our data clearly indicate that in the early 2000s the interest in complex networks, triggered by the works by Watts and Strotgatz (1998) and Barabási and Albert (1999), fueled the search for more efficient and tailored community detection algorithms.

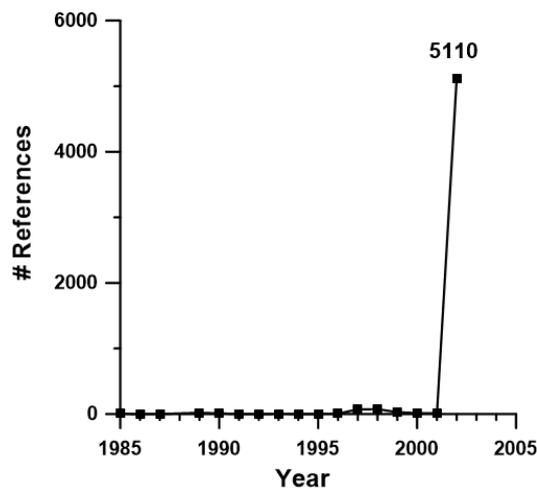

**FIG. 3** Number of cumulative references on community detection on the period 1985-2002. The squares indicate experimental data.

**RQ2.** *What is the relative weight of the different community detection methods?*

Answering this question implies to quantify the relative use, and state of development, of the different methods. In this form, we can identify over and under studied areas, an important point for determining prospective research topics.

To analyze the use of community detection methods, we need, first of all, a classification. However, no single classification does exist. Here, we use as a guide the excellent reviews by Fortunato (2010), Scheaffer (2007), Coscia et al. (2012), Xie el al. (2013) and Barabási (2016). Thus, in this work, we classify the basic community detection methods as:

- Hierarchical clustering methods: designed to reveal, specifically, the multilevel, nested, structure of a network. As canonical examples, we have the divisive Girvan-Newman (Girvan & Newman, 2002) and the agglomerative Ravasz et al. (Ravasz et al., 2002) algorithms.

- Modularity based methods: those making use of the concept of modularity introduced by Newman and Girvan (2004) as a clustering quality index to maximize. As an example, we have the greedy algorithm introduced by Newman (2004) and the also greedy Louvain algorithm (Blondel et al., 2008).

- Overlapping communities methods: which assign nodes to more than one community, such as the CFinder clique percolation algorithm (Palla et al., 2005). In this group, it is interesting to highlight the so-called fuzzy methods. They represent a generalization of the overlapping concept. Here, each node belongs, in different proportion, to each network community. The most known algorithm in this context is fuzzy c-means developed by Dunn (1973) and Bezdek (1981). Fuzzy methods are considered as an independent category in this work.

- Partitional clustering methods: here, the number of clusters is defined a priori, such as the k-means method (Macqueen, 1967).

- Spectral methods: based in the eigenvalues of some similarity matrix defined among the nodes of the network or the Laplacian matrix of the associated graph, such as the method by Donetty and Muñoz (2004).

- Dynamic methods: based on the simulation of processes in the network, such as spin coupling (Reichardt & Bornholdt, 2004), random walks (Zhou, 2003), or synchronization (Boccaletti et al., 2007).

For each of the above categories, we select the references that explicitly consider the category or specific methods within it. To such an end, we use an iterative procedure. First, we define search strings including the name of the classification categories and methods described above. From the results of this initial search, we use a "keywording" approach including in the search strings the concepts and methods found in the results datasets. In this form, we select the search strings collected in Table 2 for each of the seven considered method categories. Since not all the references include the desired information in the metadata, only a subset of the whole dataset is obtained. However, it seems a fair assumption that jobs where the type of method is not considered in its metadata tend to be applications of the most common ones. So, in the final results, the relative weight of the most common methods is probably even higher than the one obtained in the study.

**TABLE 2**. Search strings used for the different kinds of community detection methods.

| Kind of method | Search string |
|---|---|
| Hierarchical | (hierarchical cluster*) OR (hierarchical partition*) OR (Girvan-Newman) OR Ravasz OR dendogram* OR (agglomerative algorithm*) OR (divisive algorithm*) |
| Modularity | modularity AND (greedy OR optimiz* OR (simulated annealing) OR (genetic algorithm*) OR Louvain) |
| Overlapping | (overlapping communit*) OR CFinder OR (fuzzy communit*) OR (fuzzy cluster*) OR c-mean |
| Fuzzy | (fuzzy communit*) OR (fuzzy cluster*) OR c-mean |
| Partitional | (partitional cluster*) OR (k-mean*) OR (k-cluster*) OR (k-center*) OR (k-median) OR Lloyd |
| Spectral | (spectral cluster*) OR (spectral partition*) OR (spectral method*) OR laplacian OR eigenvector* |
| Dynamic | dynamic AND (Potts OR (random walk*) OR Markov OR synchronization) |

The search strings are applied to the references dataset. In this form, we determine the number of references specifically associated with each method category. The results are collected in a radar plot, see Figure 4. The first fact to remark is that only 1668 references consider explicitly in their metadata the community detection method used. Since it is unfeasible to review each of the remaining 3178 papers, we resort to statistics to determine the representativeness of our selection. Thus, applying the concept of sample size (Moore et al., 2014), we determine that a sample of 1668 elements out of 4846 corresponds to a confidence level of 99% with a confidence interval of 2.56%. So, on statistical grounds, we can consider our results to be representative enough to draw conclusions on the actual use of the different community detection methods.

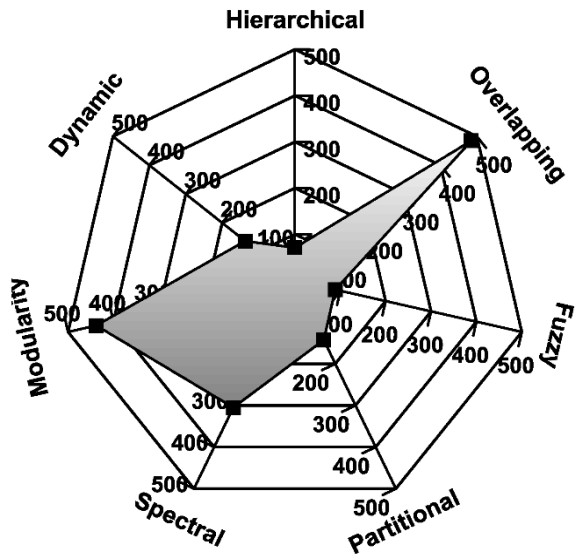

**FIG. 4** Radar plot showing the relative weight of the different kinds of community detection methods. The values indicate total number of references explicitly dealing with each kind of method.

Figure 4 shows that the category with the highest number of associated references, 485, is overlapping clustering. Modularity methods follow closely, with 438 references. These data suggest that these two categories represent interesting and active fields of study and application. In addition, as commented above, the relative importance of these methods, being the most common, is probably higher. On the other hand, the least represented category corresponds to hierarchical methods, with 71 references. However, it is interesting to remark that the concept of modularity is usually applied to determine community hierarchies in networks. So, from the application point of view, determination of hierarchical structures in networks represents the most common kind of studies (509 references considering together hierarchical and modularity methods). After pure hierarchical methods, the least represented category is fuzzy clustering, with 89 references. This last is a subset of overlapping clustering. Therefore, this is the most understudied type of methods despite the interest in overlapping clustering techniques.

**RQ3.** *What is the relative weight of the different computational models used in community detection methods/studies?*

Another aspect of interest for orientation of new research lines is the kind of computational models used, in practice, with community detection algorithms. In particular, we are interested in distinguishing the jobs that apply parallel and distributed computational systems from those using a sequential approach.

Thus, we define the search strings collected in Table 3. The first identifies all the references dealing with parallel or distributed computing in their metadata. The second locates the references specifically associated to the use of distributed computational systems. The difference corresponds to jobs focusing in some kind of parallel approach that do not use a distributed system approach.

**TABLE 3**. Search strings defined for retrieving references dealing explicitly with parallel and distributed computational environments.

| Environment | Search string |
| --- | --- |
| Non-sequential | *parallel* OR *(\*thread\*)* OR *(multiproc\*)* OR *(distributed proc\*)* OR *(distributed comput\*)* |
| Distributed | *(distributed proc\*)* OR *(distributed comput\*)* |

The first interesting result is that only 295 references, out of 4846, deal with some kind of parallel and/or distributed approach. Thus, sequential computing is by far the most used computational model in the implementation and use of community detection algorithms. So, exploiting the capabilities of current parallel/distributed systems in the present context is still a pending matter. A more detailed view can be obtained by representing publication frequencies per year. Thus, using our own software, we organize the 295 references, obtained with the search strings in Table 3, on a time scale. The results are shown in Figure 5 (where we have excluded 2016 since only partial data are available). We observe that from 1999 to 2002 no more than one job per year appears. However, since 2003 there is a steady increase of the number of references per year. As previously, we describe the time dependence using equation (1). Thus, we use the data between years 2003 and 2015, expanding the series around 2003. As before, we limit the polynomial to fourth degree and apply the stepwise selection procedure to include only the most statistically significant terms (Pekoz, 2015). The resulting polynomial is shown in equation (3),

$$n_r = -0.45 + 0.35\,(t - 2003)^2 \qquad (3)$$

with a coefficient of determination $r^2 = 0.975$ and standard deviation $\sigma = 2.87$ references.

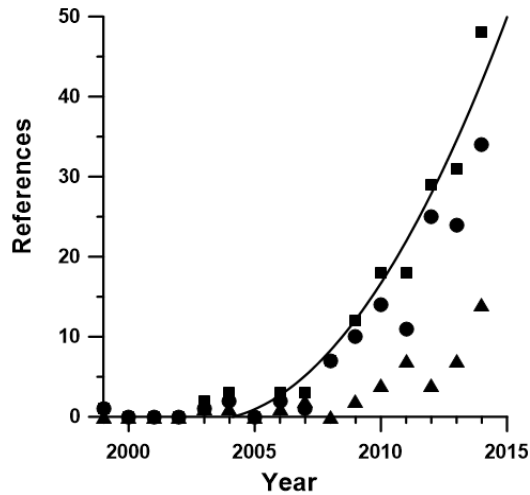

**FIG. 5** Number of references explicitly considering parallel or distributed computational environments as a function of time. Triangles and circles correspond to references dealing with distributed or parallel systems, respectively. The squares indicate total number of references. The continuous line is the regression function described in the text.

As in the case of the number of references on community detection per year, the existence of a positive increasing gradient expresses the existence of a growing interest in the topic. The change in tendency since 2003 can be explained by the so-called concurrency revolution (Díaz et al., 2012). This is the switch, around 2003, in the microprocessor industry to the multicore model. This change triggered a renewed interest in the application of the parallel computing model along all the scientific and technological fields (Díaz et al., 2012). It speaks in favor of the quality of our data that they are able to show this change of computing paradigm.

# Conclusions

In this work, we propose a bibliometric approach to Systematic Mapping Studies in order to determine, from current primary studies, the evolution, perspectives, and research opportunities of specific topics. As a case study, the approach is applied to analyze and interpret the large number of existing studies on community detection methods in complex networks. Thus, by considering the variation of the number of references over time, we find that the number of studies in the field continues to increase at a steady pace. On the grounds Rogers' diffusion of innovations theory, the large positive slope of the curve obtained by fitting the experimental data shows that, at present, the field is very active and far from exhausted.

On the other hand, analysis of the use of the different community detection methods provides the relative weight of each of them. Therefore, we find that overlapping communities methods are the most numerous. In addition, our results show that the most used kind of studies focuses on hierarchies of communities. Apart from that, the most understudied group of methods relate to fuzzy clustering despite being a subset of the popular overlapping methods category.

Finally, we consider the relative weight of the different computational models used in community detection studies. We find that only a small number of works involve specifically parallel or distributed computing. Thus, the sequential model is the most used one. In particular, the least used approach is distributed computing despite the current availability of Grid and Cloud systems. Therefore, harnessing the power of parallel, and more specifically distributed, computational systems to address large and complex community detection problems is a field still to explore.